\documentclass[pre,twocolumn,superscriptaddress,preprintnumbers,amsmath,amssymb]{revtex4-1}

\usepackage{epsfig}
\usepackage{amsmath}
\usepackage{amssymb}
\usepackage{amsfonts}
\usepackage{mathptmx}
\usepackage{dcolumn}
\usepackage{eucal}
\usepackage{bm}
\usepackage{color}
\usepackage[colorlinks,linkcolor=blue,citecolor=blue]{hyperref}
\usepackage{epstopdf}
\usepackage{bbold}
\usepackage{url}
\usepackage{enumerate}
\usepackage[table,xcdraw]{xcolor}
\usepackage{theorem}

\usepackage{dsfont}

\def\be{\begin{eqnarray}}
\def\ee{\end{eqnarray}}

\newcommand{\htot}[1]{H(z_{#1};\lambda_{#1})}
\newcommand{\hsyst}[1]{H_s(x_{#1};\lambda_{#1})}
\newcommand{\htotr}[1]{H(z^{*}_{#1};\lambda_{#1})}

\newcommand{\hbath}[1]{H_b(y_{#1})}
\newcommand{\vint}[1]{V_{int}(z_{#1})}
\newcommand{\hmf}[1]{\tilde{H}_s(x_{#1};\lambda_{#1})}

\newcommand{\rhoeq}[1]{\rho^{eq}(z_{#1};#1)}
\newcommand{\rhoeqs}[1]{\tilde{\rho}^{eq}_s(x_{#1};#1)}
\newcommand{\avbath}[1]{\langle e^{-\beta\vint{#1}} \rangle_b^{eq}}
\newcommand{\intsb}[1]{\int dz_{#1} }
\newcommand{\intb}[1]{\int dy_{#1}  }
\newcommand{\intsyst}[1]{\int dx_{#1}  }

\newcommand{\lin}{\text{ln}}
\newcommand{\pbeta}{\partial_{\beta}}

\newcommand{\average}[1]{\big\langle #1 \big\rangle}
\newcommand{\averages}[1]{\big\langle #1 \big\rangle_{s}}

\begin{document}

\title{Entropy production and time-asymmetry in the presence of strong interactions}

\author{H. J. D. Miller}
\email{hm419@exeter.ac.uk}
\author{J. Anders}
\email[]{janet@qipc.org}
\affiliation{Department of Physics and Astronomy, University of Exeter, Stocker Road, EX4 4QL, United Kingdom.}

\date{\today}

\begin{abstract}

It is known that the equilibrium properties of open classical systems that are strongly coupled to a heat bath are described by a set of thermodynamic potentials related to the system's Hamiltonian of mean force. 
By adapting this framework to a more general class of non-equilibrium states, we show that the equilibrium properties of the bath can be well-defined, even when the system is arbitrarily far from equilibrium and correlated with the bath. 
These states, which retain a notion of temperature, take the form of conditional equilibrium distributions.
For out-of-equilibrium processes we show that the average entropy production quantifies the extent to which the system-bath state is driven away from the conditional equilibrium distribution.
In addition, we show that the stochastic entropy production satisfies a generalised Crooks relation and can be used to quantify time-asymmetry of correlated non-equilibrium processes. 
These results naturally extend the familiar properties of entropy production in weakly-coupled systems to the strong coupling regime. 
Experimental measurements of the entropy production at strong coupling could be pursued using optomechanics or trapped ion systems, which allow strong coupling to be engineered. 


\end{abstract}

\date{\today}

\maketitle

\textit{Introduction.} The central goal of stochastic thermodynamics is to provide a microscopic description of entropy production at the level of the individual trajectories traced out by the system as it is driven away from equilibrium \cite{Sekimoto1998,Seifert2005,Seifert2008,Seifert2012}. Current technology now provides us with increased control over mechanically manipulated bio-molecules and nano-systems, with examples including single molecule RNA unfolding experiments \cite{Collin2005}, the manipulation of light-levitated nanospheres \cite{Millen2014} and control over trapped-ion systems \cite{an2015}. As the system size is scaled down, microscopic fluctuations in entropy become appreciable and must be understood in order to optimise the thermodynamic performance of machines and devices operating at the nanoscale \cite{Seifert2011}. On a more fundamental level entropy production provides us with a quantitative description of change and irreversibility in nature, and its average increase places restrictions on allowed state transformations in accordance with the second law of thermodynamics \cite{Clausius1865,Landau1958}. More refined statements about the nature of entropy production are given by the fluctuation theorems \cite{Crooks,Jarzynski1997d,Hatano2001b,Seifert2005,Kawai2007a,Jarzynski2011b}, and provide universal insight into the breaking of time-reversal symmetry in a wide variety of physical systems \cite{Collin2005,Feng2008a,Parrondo2009,Saira2012a,Brown2016a}. 

Standard analysis of entropy production in open systems, both quantum and classical, centres on an assumption that the system \textit{weakly} interacts with a thermal bath \cite{Kramers1940,Talkner2009,Deffner2011a,Seifert2012}. The benefit of this assumption is that it provides an unambiguous notion of stochastic heat, since neglecting energetic contributions from the interaction provides a clear division between the energy of the system and the bath. While the weak coupling assumption can be physically justified in macroscopic systems, the thermodynamic behaviour of small-scale systems may be strongly influenced by a non-negligible interaction with their environment \cite{Jarzynski2016}. Thus it is of paramount importance to explore extended notions of entropy production within the strong coupling regime, which will be the subject of this paper. 

The extension of thermodynamics to the strong coupling regime has been the subject of recent debate in the context of both classical \cite{Gelin2009a,Seifert2011,Talkner2016b,Jarzynski2016,Seifert2016} and quantum systems \cite{Jarzynski2004,Ingold2008,Esposito2010a,Hilt2011a,Carrega2016b,Philbin2016,Strasberg2016a}. The central question revolves around the identification of thermodynamic potentials for the system at both the stochastic and ensemble level. An elegant solution to this problem, originally dating back to Kirkwood in 1935 \cite{Kirkwood1935}, is to replace the isolated Hamiltonian of the system with an effective Hamiltonian that takes into account the non-negligible interaction and temperature of the environment. This allows one to define an \textit{effective} internal energy, free energy and entropy for the system at equilibrium \cite{Gelin2009a}. The operational significance of this approach is exemplified by the fact that the work done to drive a strongly interacting system through a series of equilibrium states is precisely given by $\average{W}=\Delta \tilde{F}_s$, where $\Delta \tilde{F}_s$ is the change in the \textit{effective} equilibrium free energy \cite{Jarzynski2004}, which is defined later in~(\ref{eq:thermopot}). 

Furthermore, recent efforts have extended the applicability of this formalism to stochastic, non-equilibrium thermodynamics \cite{Talkner2016b,Jarzynski2016,Seifert2016}. In particular, Seifert has proposed a definition of stochastic entropy production derived from a set of fluctuating thermodynamic potentials associated with the system's Hamiltonian of mean force \cite{Seifert2016}. In this paper we lend support to this approach by deriving an exact expression for the average entropy production in general non-equilibrium processes, and subsequently establish the second law of thermodynamics valid at arbitrary interaction strengths. Importantly it is shown that our expression converges to previously derived formulas in the limit of weak coupling \cite{Esposito2010a,Reeb2014a}. In order to consider the thermodynamics of systems operating away from equilibrium, we introduce a class $\mathcal{D}_\beta$ of system-bath configurations in which the equilibrium properties of the bath are retained even if correlated with an arbitrary state of the system that is out of equilibrium. As a result, entropy production is shown to increase as a result of the system and bath being driven away from configurations in $\mathcal{D}_\beta$. Furthermore, it is shown that the full statistics of stochastic entropy production obey a generalised Crooks-like fluctuation relation \cite{Crooks}, which provides a relationship between the time-asymmetry of non-equilibrium dynamics and the average entropy production.

We will begin by considering an open classical system coupled to a heat bath with a time-dependent Hamiltonian
\be\label{eq:ham}
	\htot{t}=\hsyst{t}+\hbath{t}+\vint{t},
\ee
where $\lambda_t$ is a time-dependent control parameter attributed to the system Hamiltonian alone, $\vint{t}$ governs the interaction between system and bath and $z_t=(x_t,y_t)$ describes a point in the collective phase space at some point in time $t$ with $x$ and $y$ labelling the system and bath degrees of freedom respectively. Let us first consider the equilibrium thermodynamics of the total system and assume a canonical distribution at inverse temperature $\beta$
\be\label{eq:thermaleq}
	\rhoeq{t}=\frac{e^{-\beta\htot{t}}}{Z(\lambda_{t})},
\ee
where $Z(\lambda_{t})=\intsb{t} \ e^{-\beta\htot{t}}$ is the partition function of the total system-bath. In standard thermodynamics one assumes that the interaction strength is sufficiently weak, $\beta\vint{t} \ll 1$, such that the total canonical state factorises into two uncorrelated canonical distributions for the system and bath respectively. In this case additive thermodynamic potentials can be assigned to both system and bath via their local equilibrium distributions. 

However, when $\vint{t}$ is non-negligible it is not immediately clear how to assign a set of thermodynamic potentials to the system. A way to solve this problem is to introduce the Hamiltonian of mean force \cite{Kirkwood1935,Jarzynski2004,Gelin2009a,Campisi2009a,Hilt2011,Philbin2016,Talkner2016b,Seifert2016,Jarzynski2016}, defined by
\be\label{eq:hmf}
	\hmf{t}:=\hsyst{t}-\frac{1}{\beta}\lin \ \avbath{t},
\ee
which acts as an effective Hamiltonian for the system that takes into account the non-negligible interaction term. Here $\langle f(z_t) \rangle_b^{eq}=\intb{t} \ f(z_t) \ e^{-\beta (\hbath{t} -F^{eq}_b)} $ denotes an average of arbitrary function $f(z_t)$ with respect to an isolated bath, and $F^{eq}_b$ is the corresponding equilibrium free energy of the isolated bath. By averaging over the bath degrees of freedom in the canonical distribution~(\ref{eq:thermaleq}), the system distribution can be expressed in an effective equilibrium state with respect to $\hmf{t}$ given by
\be\label{eq:thermaleqs}
	\rhoeqs{t}&=&\frac{e^{-\beta \hmf{t}}}{\tilde{Z}_s(\lambda_t)}, \ \ \ \  \tilde{Z}_s(\lambda_t)=\intsyst{t} \ e^{-\beta \hmf{t}}.
\ee
As was shown in \cite{Gelin2009a}, the partition function $\tilde{Z}_s(\lambda_t)$ can be used to obtain a set of thermodynamic potentials for the system through the standard formulas for free energy, internal energy and entropy:
\be\label{eq:thermopot}
	\nonumber&\tilde{F}^{eq}_s(\lambda_t)&=-\frac{1}{\beta}\lin \ \tilde{Z}_s(\lambda_t),\\
	\nonumber&\tilde{U}^{eq}_s(\lambda_t)&=-\partial_{\beta} \lin \ \tilde{Z}_s(\lambda_t), \\
	&\tilde{S}^{eq}_s(\lambda_t)&=\beta\big[\tilde{U}^{eq}_s(\lambda_t)-\tilde{F}^{eq}_s(\lambda_t)\big].
\ee
It is well known that these thermodynamic potentials are additive with respect to the bare environment \cite{Jarzynski2004,Gelin2009a,Philbin2016}. For example, the total thermodynamic entropy of the system-bath can be split into $S_{tot}^{eq}(\lambda_t)=\tilde{S}^{eq}_s(\lambda_t)+S^{eq}_b$, where $S^{eq}_b$ is the entropy of the isolated canonical bath. The same additivity holds for the internal energy and free energy, implying that the presence of the interaction leaves the equilibrium properties of the bath unchanged. Instead, the influence of the interaction is attributed to the equilibrium properties of the system alone \cite{Ford1985}. Intuitively this makes sense if one considers the environment as macroscopic relative to the microscopic size of the system. 

\textit{Non-equilibrium potentials.} While the thermodynamic potentials in~(\ref{eq:thermopot}) are well defined at equilibrium, recent efforts have attempted to extend the definitions of~(\ref{eq:thermopot}) to the case where the system is no longer in an effective equilibrium state \cite{Talkner2016b,Seifert2016,Jarzynski2016}. This can be achieved by first noting that the equilibrium internal energy can be expressed as $\tilde{U}^{eq}_s(\lambda_t)=\averages{\pbeta\big[\beta\hmf{t}\big]}^{eq}$ where $\averages{..}^{eq}$ denotes an average with respect to the effective equilibrium state~(\ref{eq:thermaleqs}). Similarly one finds $\tilde{S}^{eq}_s(\lambda_t)=-\averages{\lin \ \rhoeqs{t}}^{eq}+\beta^2 \averages{\pbeta\hmf{t}}^{eq}$. These quantities can be interpreted as equilibrium averages over a set of \textit{fluctuating} thermodynamic potentials appearing inside the brackets $\averages{..}^{eq}$. Following the general assumptions of stochastic thermodynamics \cite{Seifert2012}, we now assume that these fluctuating potentials describe the internal energy, entropy and free energy for states arbitrarily far from equilibrium, and are given respectively by \cite{Talkner2016b,Seifert2016,Jarzynski2016}
\be\label{eq:neqpot}
	\nonumber&\tilde{u}_s(x_t;\lambda_t)&:=\pbeta\big[\beta\hmf{t}\big],\\
	\nonumber&\tilde{s}_s(x_t;\lambda_t)&:=-\lin \ \rho_s(x_t;t)+\beta^2 \pbeta\hmf{t}, \\
	&\tilde{f}_s(x_t;\lambda_t)&:=\tilde{u}_s(x_t;\lambda_t)-\beta^{-1}\tilde{s}_s(x_t;\lambda_t).	
\ee 
While these definitions are initially taken as postulates, it will ultimately be shown that these fluctuating potentials can be connected into a consistent thermodynamic framework. We will denote the average non-equilibrium internal energy by $\tilde{U}_s(\lambda_t;t)=\averages{\tilde{u}_s(x_t;\lambda_t)}$, with $\averages{..}=\intsyst{t} \ \rho_s(x_t;t)(..)$ now an average with respect to a general non-equilibrium state of the system. Similarly the average entropy will be denoted by $\tilde{S}_s(\lambda_t;t)=\averages{\tilde{s}_s(x_t;\lambda_t)}$ and average free energy by $\tilde{F}_s(\lambda_t;t)=\averages{\tilde{f}_s(x_t;\lambda_t)}$. From~(\ref{eq:neqpot}) one sees that the non-equilibrium entropy at strong coupling involves a contribution from the Gibbs-Shannon entropy alongside a second term $\beta^2\averages{\pbeta\hmf{t}}$ that has previously been identified as an intrinsic entropy in the context of small-scale molecular motors \cite{Seifert2011}. 

As one might expect, these potentials will not generally be additive for a given system-bath distribution, unlike the equilibrium counterparts~(\ref{eq:thermopot}). However, let us consider a particular class $\sigma(z_t;t)\in\mathcal{D}_{\beta}$ of distributions defined by
\be\label{eq:stat}
	\sigma(z_t;t)=\rho_s(x_t;t)\rho^{eq}_b(y_t|x_t;\lambda_t),
\ee
where we place no restriction on the system configuration and 
\be
	\rho^{eq}_b(y_t|x_t;\lambda_t)=\frac{\rho^{eq}(z_t;\lambda_t)}{\intb{t}\rho^{eq}(z_t;\lambda_t)},
\ee
is the \emph{equilibrium} conditional probability for bath micro-state $y_t$ given a particular micro-state of the system $x_t$, obtained through application of Bayes' theorem. The class of states $\mathcal{D}_{\beta}$ has previously been introduced in \cite{Talkner2016b} and referred to as the \textit{stationary preparation class}, which describes a conditional equilibrium state on the bath. In this case for any micro-state selected from the system the resulting conditional statistics of the bath are equivalent to that of the total canonical state~(\ref{eq:thermaleq}). For this particular class of states, which of course includes the canonical state~(\ref{eq:thermaleq}), one still has a well defined notion of temperature attributed to a thermal environment. This is exemplified by a \textit{generalised} additive relationship between the thermodynamic potentials, which we prove in Appendix~\ref{app:1}. Taking the state $\sigma(z_t;t)\in\mathcal{D}_\beta$, let us denote $U_{tot}(\lambda_t;t)=\average{\htot{t}}$ as the internal energy of $\sigma(z_t;t)$, $S_{tot}(\lambda_t;t)=-\average{\lin \ \sigma(z_t;t)}$ the Gibbs-Shannon entropy and $F_{tot}(\lambda_t;t)=U_{tot}(\lambda_t;t)-\beta^{-1}S_{tot}(\lambda_t;t)$ as the free energy. Then the following additive property holds:
\be\label{eq:extensive}
	 \chi_{tot}(\lambda_t;t)=\tilde{\chi}_s(\lambda_t;t)+\chi^{eq}_{b},
\ee
where $\chi\in\lbrace F,S,U\rbrace$. This implies that the equilibrium properties of the bath remain well-defined relative to the arbitrary state of the system, even in the presence of correlations due to strong interaction. Ultimately this additive property of the class $\mathcal{D}_\beta$ will allow us to maintain a notion of temperature for states driven away from equilibrium, and will allow us to derive the second law of thermodynamics in this framework. 

\textit{Entropy production.} We will now consider a general non-equilibrium (NEQ) process operating at an arbitrarily large coupling strength and derive an exact expression for the entropy production. The NEQ process is realised over a time interval $[t_0,t]$ by varying the Hamiltonian through a parameter change $\lambda_t$ with initial and final settings denoted by $\lambda_0$ and $\lambda_t$ respectively. We make two assumptions about this process:
\begin{enumerate}[(i)]
\item At initial time $t_0$ the system-bath is in a conditional equilibrium state $\sigma(z_0;t_0)\in\mathcal{D}_\beta$, with $\rho_s(x_0;t_0)$ specifying an initial arbitrary state for the system.
\item The total system-bath undergoes closed evolution during the time interval $[t_0,t]$ governed by Liouville's equation
\be\label{eq:evo}
	\partial_t\rho(z_t;t)=\mathcal{L}\big[\rho(z_t;t)],
\ee
where $\mathcal{L}[(..)]$ is the corresponding Liouvillian resulting from the change in the Hamiltonian~(\ref{eq:ham}) over time. The resulting final state is specified by $\rho(z_t ;t)$ with final system configuration $\rho_s (x_t;t)=\intb{t} \ \rho(z_t;t)$. 
\end{enumerate}
Assumption (i) is necessary in order to have a well-defined notion of both temperature and the Hamiltonian of mean force~(\ref{eq:hmf}) prior to the NEQ process. Assumption (ii) ensures that we account for all exchanges of heat and work between the system and the bath. No restrictions are imposed on the final configuration of the system, and we denote the transformation by $\rho_s(x_0;t_0)\rightarrow \rho_s(x_t;t)$. Following the approaches taken in \cite{Talkner2016b,Seifert2016,Jarzynski2016} we can use the fluctuating potentials in~(\ref{eq:neqpot}) to define the fluctuating heat dissipated from the system into the bath up to time $t$ as
\be\label{eq:heat2}
	\tilde{Q}(z_t;t):=\tilde{u}_s(x_0;\lambda_0)-\tilde{u}_s(x_t;\lambda_t)+\int^{t}_{t_0}dt \ \partial_t\tilde{u}_s(x_t;\lambda_t),
\ee
which represents the sum of work done during the process and the decrease in internal energy of the system, in accordance with the first law of thermodynamics. Note that $\tilde{Q}(z_t;t)=\tilde{Q}(z_t[z_0];t)$ is implicitly written as a function of the initial phase space point $z_0$  because the evolution of point $x_t$ depends on the deterministic evolution of the collective phase space for the system and bath, denoted by the transformation $z_0\rightarrow z_t[z_0]$. However, the RHS of~(\ref{eq:heat2}) indicates that the heat can be determined by monitoring the system degrees of freedom alone along a specific trajectory. If we take into account the full evolution of the system-bath, it is straightforward to show that the average dissipated heat is given by
\be\label{eq:heat}
	\average{\tilde{Q}(t)}=U_{tot}(\lambda_t;t)&-&\tilde{U}_s(\lambda_t;t)-U^{eq}_b,
\ee
which follows from~(\ref{eq:extensive}) combined with initial condition (i), along with the fact that the integral in~(\ref{eq:heat2}) is equivalent to the difference in total energy, $\htot{t}-\htot{0}$. As noted by Seifert, one can introduce a definition of fluctuating entropy production as the sum of dissipated heat and change in the fluctuating entropy of the system \cite{Seifert2016};
\be\label{eq:entprod}
	\Sigma(z_t;t):=\tilde{s}_s(x_t;\lambda_t)-\tilde{s}_s(x_0;\lambda_0)+\beta \tilde{Q}(z_t;t).
\ee
For the definition~(\ref{eq:entprod}) to be a physically relevant candidate for entropy production then it should not be negative on average, in accordance with the second law of thermodynamics. This brings us to the main result of the paper.

\textit{Main result.} Assuming the total system-bath undergoes the NEQ process specified by assumptions (i) and (ii), then the average entropy production up to time $t$ is given by
\be\label{eq:land}
	\average{\Sigma(t)}=S[\rho(z_t ;t)||\sigma(z_t;t)],	
\ee
where 
\be	
	\nonumber S[\rho(z_t ;t)||\sigma(z_t;t)]=\intsb{t} \rho(z_t ;t) \ \lin \ \bigg[\frac{\rho(z_t ;t)}{\sigma(z_t;t)}\bigg],
\ee
is the Kullback-Leibler divergence between the final system-bath configuration and the corresponding conditional equilibrium state $\sigma(z_t;t)=\rho_s(x_t;t)\rho^{eq}_b(y_t|x_t;\lambda_t)\in\mathcal{D}_{\beta}$. This is the central result of the paper and the proof of~(\ref{eq:land}) is provided in Appendix~\ref{app:2}. By~(\ref{eq:land}) and the positivity of the Kullback-Leibler divergence, one has $\average{\Sigma(t)}\geq 0$ as desired. From the definition of entropy production in~(\ref{eq:entprod}) one obtains a form of the Clausius inequality valid for arbitrary coupling strengths which becomes 
\be\label{eq:land2}
	\beta\average{\tilde{Q}(t)}\geq \tilde{S}_s(\lambda_0;t_0)-\tilde{S}_s(\lambda_t;t).
\ee
Perhaps surprisingly, the Clausius inequality derived here within the strong coupling regime suggests that the change in Gibbs-Shannon entropy is generally insufficient to bound the minimum heat dissipated into the bath during a non-equilibrium process. Apparent violations of the usual Clausius inequality resulting from correlations between the system and bath in the strong coupling regime have previously been observed in \cite{Allahverdyan2000,Hilt2011a}. However, we see here that a consistent form of the second law can be recovered if one considers the additional temperature-dependent terms appearing in the fluctuating entropy~(\ref{eq:neqpot}).

\begin{figure}[t]
\hspace*{-0.4cm} 
\includegraphics[scale=0.37]{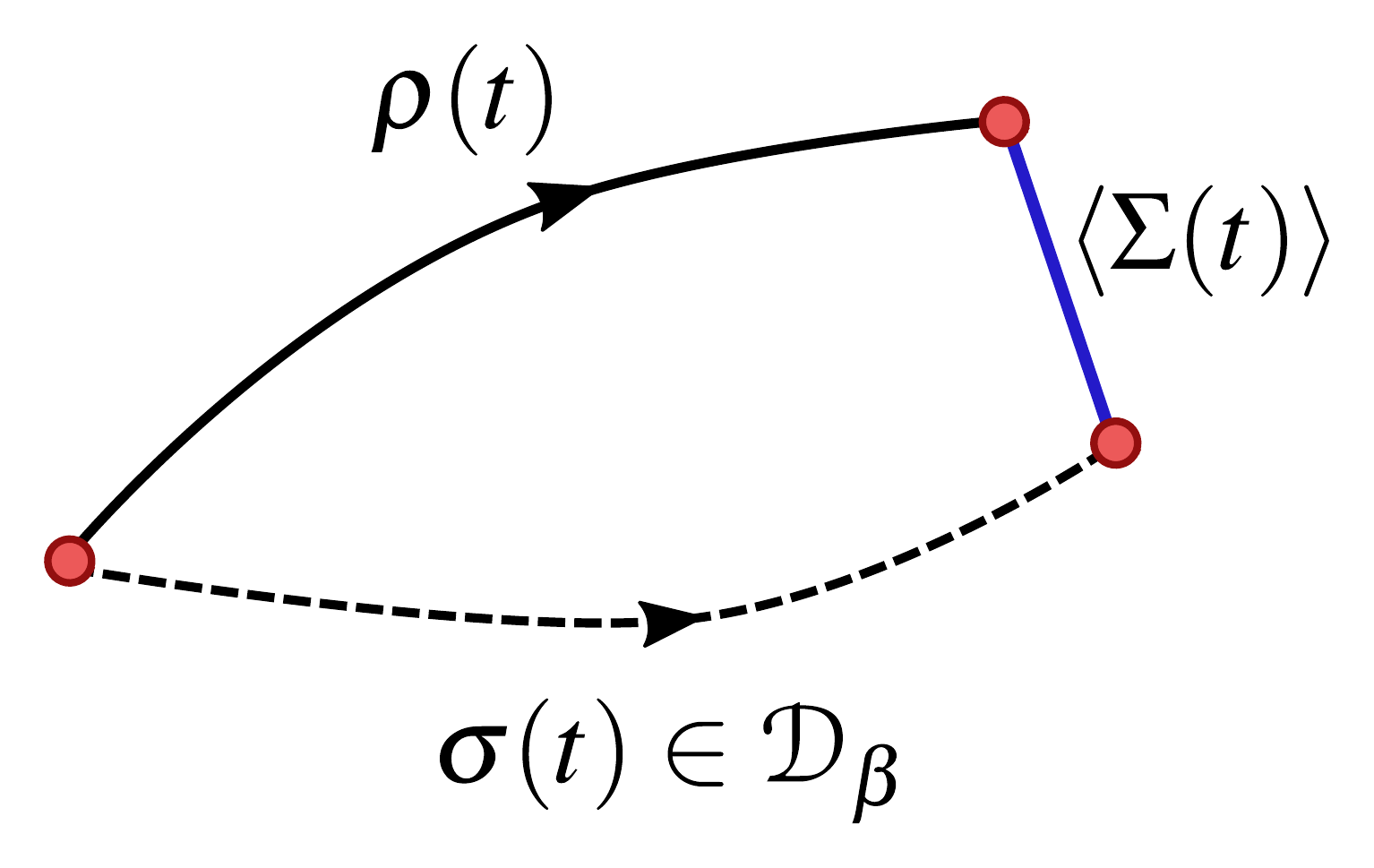}
\caption{\label{fig:lag}Schematic representation of the equality~(\ref{eq:land}). The solid line represents the actual process given by the evolving distribution $\rho(t)=\rho(z_t ;t)$ whilst the dashed line represents a fictitious quasi-static process in which the system-bath distribution stays in the conditional equilibrium state $\sigma(t)=\sigma(z_t;t)\in\mathcal{D}_\beta$. The non-negative entropy production then quantifies the extent to which the system and bath are driven away from $\sigma(t)$, represented here as the distance of the blue line.}
\end{figure}

According to Stein's lemma \cite{cover2012elements}, the divergence appearing in~(\ref{eq:land}) can be interpreted as a measure of distinguishability between the final distribution and the corresponding conditional equilibrium state $\sigma(z_t;t)\in\mathcal{D}_\beta$. Thus the further the final state is driven away from the uniquely defined $\sigma(z_t;t)\in\mathcal{D}_\beta$, the greater the amount of entropy production after the process. If the dynamics governed by~(\ref{eq:evo}) are such that the total system-bath remains in the corresponding conditional equilibrium state in $\mathcal{D}_{\beta}$, the bound in~(\ref{eq:land2}) can be saturated at any given time $t$. However, in this situation the dissipated heat and entropy change are simultaneously zero; $\beta\average{\tilde{Q}(t)}=\Delta \tilde{S}_s=0$. The expression~(\ref{eq:land}) can be interpreted as a generalisation of a phenomenon known as lag encountered in closed/weakly-coupled thermodynamic systems \cite{Vaikuntanathan2009}. In particular, we see that the entropy production quantifies the extent to which the configuration of the system-bath \textit{lags} behind a hypothetical quasi-static process in which the configuration remains in the evolving conditional equilibrium state, $\sigma(z_t;t)\in\mathcal{D}_\beta$. An illustration of this effect is represented in Figure~\ref{fig:lag}.

Result~(\ref{eq:land}) is consistent with previously derived expressions for average entropy production when the weak-coupling limit is taken. If one assumes $\beta\vint{t} \ll 1$ then the Hamiltonian of mean force~(\ref{eq:hmf}) reduces to the system Hamiltonian $\hsyst{t}$ independent of temperature. As expected the heat becomes $\average{\tilde{Q}(t)}\approx\average{H_b(t)}-\average{H_b(t_0)}$, where $\average{H_b(t)}$ is the average energy of the isolated bath Hamiltonian evaluated with respect to the configuration of the bath at time $t$. Secondly, this also means the entropy change reduces to the change in Gibbs-Shannon entropy $\tilde{S}_s(\lambda_t;t)\approx S_s(t)= -\intsyst{t}\rho_s(x_t;t)\lin \ \rho_s(x_t;t)$. Finally, it can also be seen that the conditional equilibrium state $\sigma(z_t;t)\in\mathcal{D}_{\beta}$ reduces to a system state uncorrelated with the isolated canonical bath; $\sigma(z_t;t)\approx \rho_s(x_t;t)\rho_b^{eq}(y_t)$. By comparison with~(\ref{eq:land}), we obtain the same equality derived in \cite{Esposito2010a,Reeb2014a} which is
\be\label{eq:claus}
	\nonumber\average{\Sigma(t)}&\approx& S_s(t)-S_s(t_0)+\beta\average{H_b(t)}-\beta\average{H_b(t_0)}, \\
	&=&S[\rho(z_t ;t)||\rho_s(x_t;t)\rho_b^{eq}(y_t)].
\ee
where $\rho_b^{eq}(y_t)=e^{-\beta(\hbath{t}-F^{eq}_b)}$. It should be noted that~(\ref{eq:claus}) was originally derived for quantum systems in \cite{Esposito2010a,Reeb2014a}, though in the weak-coupling regime the result is entirely statistical-mechanical in nature and continues to hold in classical systems. 

\textit{Fluctuation theorem.} We have demonstrated that the average entropy production $\average{\Sigma(t)}$ quantifies the extent to which the total system-bath is driven away from states in $\mathcal{D}_{\beta}$. This suggests that the fluctuations in $\Sigma(z_t;t)$ can be used to quantify time-asymmetry in the dynamics of strongly coupled systems. In both weakly-coupled and closed systems, fluctuation relations can be used to indicate a breaking of time-reversal symmetry by comparing the statistics of positive entropy production for a forward trajectory versus negative entropy production along the corresponding time-reversed trajectory \cite{Crooks,Hatano2001b,Kawai2007a,Feng2008a,Gomez-Marin2008,Parrondo2009}. We will now show that the entropy production satisfies a Crooks-like fluctuation relation. Let us again suppose that we drive a system-bath configuration $\sigma(z_0;t_0)\in\mathcal{D}_{\beta}$ away from $\mathcal{D}_\beta$ by varying the control parameter $\lambda_0\rightarrow\lambda_t$, and denote the initial and final configurations of the system by $\rho_s(x_0;t_0)$ and $\rho_s(x_t;t)$ respectively. The stochastic entropy production $\Sigma(z_t;t)$ along a particular phase space trajectory fluctuates according to the sampling of the initial phase space point, and the resulting probability of occurrence can be written as follows;
\be\label{eq:entprob}
	\overrightarrow{P}(+\Sigma)=\intsb{i} \ \sigma(z_0;t_0) \ \delta[\Sigma-\Sigma(z_t;t)],
\ee
where the superscript indicates that the process moves forwards in time. To compare this with the time-reversed entropy production we need to make additional assumptions. Firstly, we require the total Hamiltonian to be time-reversal symmetric, $\htot{t}=\htotr{t}$, where $z^{*}_t$ indicates a conjugated phase space point in which momentum is reversed. Secondly, the initial and final configurations of the system are assumed to be time reversal symmetric; $\rho_s(x_0;t_0)=\rho_s(x^{*}_0;t_0)$ and $\rho_s(x_t;t)=\rho_s(x^{*}_t;t)$. By comparison with~(\ref{eq:heat2}) and~(\ref{eq:entprod}) it is straightforward to see that these conditions imply $\Sigma(z_t;t)=-\Sigma(z^{*}_t;t)$. For the time-reversed process, the initial configuration is given by $\sigma(z^{*}_t;t)=\rho_s(x^{*}_t;t)\rho^{eq}_b(y^{*}_t|x^{*}_t;\lambda_t)\in\mathcal{D}_\beta$ and the control parameter is varied from $\lambda_t\rightarrow\lambda_0$. As with~(\ref{eq:entprob}), entropy production along the reverse process has a corresponding probability of occurrence denoted by $\overleftarrow{P}(-\Sigma)$. As is proven in Appendix~\ref{app:3}, these probabilities are related by a fluctuation relation,
\be\label{eq:crooks}
	\frac{\overrightarrow{P}(+\Sigma)}{\overleftarrow{P}(-\Sigma)}=e^{+\Sigma},
\ee
implying that a positive entropy production along the forward trajectory is exponentially favoured against its time reverse. Taking the logarithm of both sides and performing an average over $\overrightarrow{P}(+\Sigma)$ yields an alternative expression for the average entropy production:
\be\label{eq:symm}
	\average{\Sigma(t)}=S\big[\overrightarrow{P}(+\Sigma)||\overleftarrow{P}(-\Sigma)\big].
\ee
Following Stein's lemma again, we see that the average entropy production also quantifies the distinguishability between statistics of the forward and reverse non-equilibrium processes respectively. By comparison with~(\ref{eq:land}), if the dynamics are such that the system and bath remain in their corresponding configuration in $\mathcal{D}_\beta$ then the LHS of~(\ref{eq:symm}) reduces to zero, implying the dynamics are completely symmetric in time as expected \cite{Feng2008a,Vaikuntanathan2009}. This solidifies our interpretation of the entropy production~(\ref{eq:entprod}) as a measure of time-asymmetry and irreversibility generalised to the strong coupling regime.  

\textit{Conclusion.} In this paper we have shown that the entropy production in a system strongly interacting with a bath demonstrates a positive increase in accordance with the second law of thermodynamics. In particular, we proved that entropy is produced when the system and bath are driven away from the conditional equilibrium distribution in $\mathcal{D}_\beta$. As we have argued, the stochastic entropy production~(\ref{eq:entprod}) is accessible through monitoring the system's path in phase space, implying that in principle a verification of our results~(\ref{eq:land}) and~(\ref{eq:crooks}) should be accessible using standard experimental techniques \cite{Collin2005,Tietz2006}. Our results may provide important modifications to Landauer's principle \cite{Reeb2014a} in the presence of strong coupling, as the change in Shannon entropy is insufficient to characterise the minimum heat dissipated into the bath as the result of information-erasure, as shown in the generalised Clausius inequality~(\ref{eq:land2}). 

\textit{Acknowledgments}: HM is supported by EPSRC through a Doctoral Training Grant. HM would like to thank Hamed Mohammady for useful discussions and a critical reading of the manuscript. JA acknowledges support from EPSRC, grant EP/M009165/1, and the Royal Society. This research was supported by the COST network MP1209 `Thermodynamics in the quantum regime'.




\bibliographystyle{apsrev4-1}
\bibliography{phd_land}

\appendix
\widetext

\section{Proof of~(\ref{eq:extensive})}\label{app:1}

In this section we prove that for any conditional equilibrium distribution $\rho(z_t;t)=\rho_s(x_t;t)\rho^{eq}_b(y_t|x_t;\lambda_t)\in\mathcal{D}_{\beta}$, the non-equilibrium potentials~(\ref{eq:neqpot}) satisfy the additive property $\chi_{tot}(\lambda_t;t)=\tilde{\chi}_s(\lambda_t;t)+\chi^{eq}_{b}$. To express $\rho(z_t;t)$ in a more useful form we use the following identity \cite{Talkner2016b}; 
\be
	\nonumber \rho^{eq}_b(y_t|x_t;\lambda_t)&=&\frac{\rho^{eq}(z_t;\lambda_t)}{\intb{t}\rho^{eq}(z_t;\lambda_t)}, \\
	&=&\frac{e^{-\beta(\hbath{t}+\vint{t})}}{\intb{t} \ e^{-\beta(\hbath{t}+\vint{t})}}.
\ee
We now note that the non-equilibrium internal energy is given by $\tilde{U}_s(\lambda_t;t)=\averages{\pbeta\big[\beta\hmf{t}\big]}$. To proceed we expand the fluctuating internal energy function $\tilde{u}_s(x_t;\lambda_t)=\pbeta\big[\beta\hmf{t}\big]$;
\be\label{eq:us}
	\nonumber \tilde{u}_s(x_t;\lambda_t)&=&\pbeta\big[\beta\hmf{t}\big], \\
	\nonumber&=& \hsyst{t}-\frac{\pbeta \avbath{t}}{\avbath{t}}, \\
	\nonumber&=&\hsyst{t}+\frac{\intb{t} \ e^{-\beta(\hbath{t}+\vint{t})}[\hbath{t}+\vint{t}]}{\intb{t} \ e^{-\beta(\hbath{t}+\vint{t})}}+\pbeta \big[e^{-\beta F^{eq}_b}\big], \\
	&=& \hsyst{t}+\intb{t} \ \rho^{eq}_b(y_t|x_t;\lambda_t)[\hbath{t}+\vint{t}]-U^{eq}_b.
\ee
Averaging both sides~(\ref{eq:us}) with respect to $\rho_s(x_t;t)$ gives
\be
	\nonumber \tilde{U}_s(\lambda_t;t)&=& \intsb{t} \ \rho_s(x_t;t)\rho^{eq}_b(y_t|x_t;\lambda_t)[\hsyst{t}+\hbath{t}+\vint{t}] -U^{eq}_b,\\
	&=&U_{tot}(\lambda_t;t) -U^{eq}_b.
\ee
Turning now to the entropy, we need to evaluate the Gibbs-Shannon entropy of the state $\rho(z_t;t)\in\mathcal{D}_{\beta}$. This can be done from the following equivalent identity;
\be\label{eq:statprep}
	\rho^{eq}_b(y_t|x_t;\lambda_t)=e^{-\beta(\htot{t}-\hmf{t}-F^{eq}_b)}.
\ee
Using this we can show the following
\be\label{eq:ss}
	\nonumber S_{tot}(\lambda_t;t)&=&-\intsb{t} \ \rho_s(x_t;t)\rho^{eq}_b(y_t|x_t;\lambda_t)[\lin \ \rho_s(x_t;t)+\lin \ \rho^{eq}_b(y_t|x_t;\lambda_t)],\\
	\nonumber &=& S_s(\lambda_t;t)-\beta F^{eq}_b+\beta\intsb{t} \ \rho_s(x_t;t)\rho^{eq}_b(y_t|x_t;\lambda_t)[\htot{t}-\hmf{t}],\\
	\nonumber &=&S_s(\lambda_t;t)-\beta (U_{tot}(\lambda_t;t)-U^{eq}_b)+S^{eq}_b-\beta\intsyst{t} \ \rho_s(x_t;t)\hmf{t}, \\
	\nonumber&=&S_s(\lambda_t;t)+\beta\tilde{U}_{s}(\lambda_t;t)-\beta\averages{\hmf{t}}+S^{eq}_b, \\
	&=&\tilde{S}_{s}(\lambda_t;t)+S^{eq}_b,
\ee
where we used $U_{tot}(\lambda_t;t)-U^{eq}_b=\tilde{U}_{s}(\lambda_t;t)$ and $\beta^{2}\averages{\pbeta\hmf{t}}=\beta\tilde{U_{s}}(\lambda_t;t)-\beta\averages{\hmf{t}}$. Finally, the last additive relation 
\be
	F_{tot}(\lambda_t;t)=\tilde{F}_s(\lambda_t;t)+F^{eq}_{b},
\ee	
follows trivially from~(\ref{eq:us}) and~(\ref{eq:ss}) together with the definition of fluctuating free energy, $\tilde{f}_s(x_t;\lambda_t)=\tilde{u}_s(x_t;\lambda_t)-\beta^{-1}\tilde{s}_s(x_t;\lambda_t)$. This concludes the proof of~(\ref{eq:extensive}).

\section{Proof of~(\ref{eq:land})}\label{app:2}

We begin by expressing the decrease in non-equilibrium entropy for the NEQ process specified by assumptions (i) and (ii) in the main text as follows;
\be\label{eq:entropy}
	\nonumber \Delta \tilde{S}_s&=& \tilde{S}_s(\lambda_0;t_0)-\tilde{S}_s(\lambda_t;t),\\
	\nonumber &=&	 S_{tot}(\lambda_0;t_0)-S^{eq}_b-\tilde{S}_s(\lambda_t;t),\\
	\nonumber&=& S_{tot}(\lambda_t;t)-S^{eq}_b-S_s(\lambda_t;t)-\beta^{2}\averages{\pbeta\hmf{t}},\\
	&=&S_{tot}(\lambda_t;t)-S^{eq}_b-S_s(\lambda_t;t)-\beta\tilde{U}_s(\lambda_t;t)+\beta\averages{\hmf{t}},
\ee
where we recall $S_s(t)= \intsyst{t}\rho_s(x_t;t)\lin \ \rho_s(x_t;t)$ represents the Gibbs-Shannon entropy of the system. In the second line we applied the additivity of the non-equilibrium entropy, according to~(\ref{eq:extensive}). This is ensured by our choice of initial conditions given by assumption (i). In the third line we used the fact that the Gibbs-Shannon entropy is invariant under closed evolution given by~(\ref{eq:evo}) \cite{Parrondo2009}. The remaining steps follow from the definitions of $\tilde{S}_s(\lambda_t;t)$ and $\tilde{U}_s(\lambda_t;t)$. 

Now we introduce the Kullback-Leibler divergence $S[\rho(z_t ;t)||\sigma(z_t;t)]$ defined in~(\ref{eq:land}). Using $\sigma(z_t;t)=\rho_s(x_t;t)\rho^{eq}_b(y_t|x_t;\lambda_t)$ according to~(\ref{eq:statprep}), the KL divergence can be evaluated as follows;
\be\label{eq:KLD}
	\nonumber S[\rho(z_t ;t)||\sigma(z_t;t)]&=&\intsb{f} \rho(z_t ;t) \ \lin \ \bigg[\frac{\rho(z_t ;t)}{\sigma(z_t;t)}\bigg] \\
	\nonumber &=& -S_{tot}(\lambda_t;t)+S_{s}(\lambda_t;t)-\intsb{f} \ \rho(z_t ;t)\lin \ \rho^{eq}_b(y_t|x_t;\lambda_t)\\
	\nonumber &=& -S_{tot}(\lambda_t;t)+S_{s}(\lambda_t;t)-\beta F^{eq}_b+\beta\average{\htot{f}}-\beta\averages{\hmf{f}}, \\
	&=&-\Delta  \tilde{S}_s+\beta[U_{tot}(\lambda_t;t)-\tilde{U}_s(\lambda_t;t)-U_b^{eq}].
\ee
where we used~(\ref{eq:entropy}) and $F^{eq}_b=U_b^{eq}-\beta^{-1}S^{eq}_b$ in the final line. By using $\tilde{U}_s(\lambda_0;t_0)=U_{tot}(\lambda_0;t_0)-U^{eq}_b$ from~(\ref{eq:extensive}), it is straightforward to see that the dissipated heat~(\ref{eq:heat}) takes the form
\be
	\nonumber\average{\tilde{Q}(t)}&=&[U_{tot}(\lambda_t;t)-\tilde{U}_s(\lambda_t;t)]-[U_{tot}(\lambda_0;t_0)-\tilde{U}_s(\lambda_0;t_0)], \\
	&=&U_{tot}(\lambda_t;t)-\tilde{U}_s(\lambda_t;t)-U_b^{eq}.
\ee
Finally, we combine~(\ref{eq:KLD}) with~(\ref{eq:heat2}) to arrive at 
\be
	S[\rho(z_t ;t)||\sigma(z_t;t)]=\beta\average{\tilde{Q}(t)}-\Delta  \tilde{S}_s=\average{\Sigma(t)},
\ee
thus concluding the proof of~(\ref{eq:land}).

\section{Proof of~(\ref{eq:crooks})}\label{app:3}

To begin, first note that the fluctuating heat~(\ref{eq:heat}) can be expressed in terms of the difference between the fluctuating total energy and fluctuating internal energy of the system;
\be
	\tilde{Q}(z_t;t)=[\htot{t}-\tilde{u}_s(x_t;\lambda_t)]-[\htot{0}-\tilde{u}_s(x_0;\lambda_0)].
\ee
Recall that the initial state for the forward process is specified by $\sigma(z_0;t_0)=\rho_s(x_0;t_0)\rho^{eq}_b(y_0|x_0;\lambda_0)$, whilst for the time-reversed process the initial configuration is given by $\sigma(z^{*}_t;t)=\rho_s(x^{*}_t;t)\rho^{eq}_b(y^{*}_t|x^{*}_t;\lambda_t)\in\mathcal{D}_\beta$. Using~(\ref{eq:statprep}) we expand the following;
\be\label{eq:ratio}
	\nonumber\lin \ \bigg[\frac{\sigma(z_0;t_0)}{\sigma(z^{*}_t;t)}\bigg]&=&\lin \ \bigg[\frac{\rho_s(x_0;t_0)\rho^{eq}_b(y_0|x_0;\lambda_0)}{\rho_s(x^{*}_t;t)\rho^{eq}_b(y^{*}_t|x^{*}_t;\lambda_t)}\bigg], \\
	\nonumber&=&\lin \ \bigg[\frac{\rho_s(x_0;t_0)}{\rho_s(x^{*}_t;t)}\bigg]-\beta\big[\htot{0}-H(z^{*}_t;\lambda_t)-\hmf{0}+\tilde{H}_s(x^{*}_{t};\lambda_{t})\big], \\
	\nonumber&=&\lin \ \bigg[\frac{\rho_s(x_0;t_0)}{\rho_s(x_t;t)}\bigg]-\beta\big[\htot{0}-H(z_t;\lambda_t)-\hmf{0}+\tilde{H}_s(x_{t};\lambda_{t})\big], \\
	\nonumber&=&\tilde{s}_s(x_t;\lambda_t)-\tilde{s}_s(x_0;\lambda_0)-\beta\big[\htot{0}-H(z_t;\lambda_t)-\hmf{0}+\tilde{H}_s(x_{t};\lambda_{t})-\beta^{2}\partial_\beta\hmf{0}+\beta^{2}\partial_\beta\hmf{t}\big], \\
	&=&\tilde{s}_s(x_t;\lambda_t)-\tilde{s}_s(x_0;\lambda_0)+\beta \tilde{Q}(z_t;t),
\ee
where we used the time-reversal symmetry assumptions for $\htot{t}$ and $\rho_s(x_t;t)$ and in the final line applied the definition~(\ref{eq:entprod}). The above equality represents a detailed balanced relation that can be used to prove~(\ref{eq:crooks}). We now evaluate the probability $\overleftarrow{P}(-\Sigma)$;
\be
	\overleftarrow{P}(-\Sigma)&=& \intsb{t}^{*} \ \sigma(z^{*}_t;t) \ \delta[\Sigma+\Sigma(z^{*}_t)], \\
	\nonumber&=& \intsb{0} \ \bigg|\frac{\partial z^{*}_t}{\partial z_0}\bigg|^{-1} \ \bigg[\frac{\sigma(z^{*}_t;t)}{\sigma(z_0;t_0)}\bigg] \ \sigma(z_0;t_0) \ \delta[\Sigma-\Sigma(z_t;t)], \\
	\nonumber&=&\intsb{0} \ e^{-\tilde{s}_s(x_t;\lambda_t)+\tilde{s}_s(x_0;\lambda_0)-\beta \tilde{Q}(z_t;t)} \ \sigma(z_0;t_0) \ \delta[\Sigma-\Sigma(z_t;t)], \\
	\nonumber&=&e^{-\Sigma}\intsb{0} \ \sigma(z_0;t_0) \ \delta[\Sigma-\Sigma(z_t;t)], \\
	&=&e^{-\Sigma} \ \overrightarrow{P}(+\Sigma),
\ee
where in the second line we performed a change of variables $z^{*}_t\rightarrow z_0$ along with $\Sigma(z_t;t)=-\Sigma(z^{*}_t;t)$, in the third line we used the fact that the Jacobian is equal to unity and~(\ref{eq:ratio}), and in the fourth line we pulled the exponential outside the integral due to the presence of the delta function. This concludes the proof of~(\ref{eq:crooks}). 

\end{document}